\documentclass[fleqn,usenatbib]{mnras}

\usepackage[T1]{fontenc}
\usepackage{ae,aecompl}
\usepackage[dvipsnames]{xcolor}
\usepackage{amssymb}

\usepackage{graphicx}	
\usepackage{amsmath}	
\usepackage{amssymb}	
\usepackage{listings}
\usepackage{soul,xcolor}

\newcommand{\msun}{M$_\odot$}
\setstcolor{red}

\title[High-$z$ galaxies from bumpy power spectrum]{Excess of high-$z$ galaxies as a test for bumpy power spectrum of density perturbations
}

\author[M.V. Tkachev, S.V. Pilipenko, E.V. Mikheeva, V.N. Lukash]{
M.V. Tkachev$^{1}$\thanks{mtkachev@asc.rssi.ru}
S.V. Pilipenko$^{1}$\thanks{spilipenko@asc.rssi.ru}
E.V. Mikheeva$^{1}$\thanks{helen@asc.rssi.ru}
V.N. Lukash$^{1}$\thanks{lukash@asc.rssi.ru}
\newauthor
\\
$^{1}$Astro Space Center of P. N. Lebedev Physical Institute of RAS, Profsojuznaya 84/32, Moscow 117997, Russia\\
}

\date{Accepted XXX. Received YYY; in original form ZZZ}

\pubyear{2023}

\begin{document}
\label{firstpage}
\pagerange{\pageref{firstpage}--\pageref{lastpage}}
\maketitle

\begin{abstract}
Modified matter power spectra with approximately Gaussian bump on sub-Mpc scales can be a result of a complex inflation. We consider five spectra with different Gaussian amplitudes $A$ and locations $k_0$ and run N-body simulations in a cube $(5 Mpc/h)^3$ at $z>8$ to reveal the halo mass functions and their evolution with redshift. We have found that the Sheth-Tormen formula provides a good approximation to a such kind of 
halo mass functions. In the considered models the dark matter halo formation starts much more earlier than in $\Lambda$CDM, which in turn can result in an earlier star formation and a nuclear activity in galaxies and can be detected and tested by, e.g., JWST. 
At $z=0$ the halo mass functions are hardly distinguishable from the standard $\Lambda$CDM, therefore the models with the bumpy spectra can be identified in observations by their excess in number of bright sources at high redshift only.

\end{abstract}

\begin{keywords}
cosmology-simulations --- dark matter --- matter power spectrum
\end{keywords}


\section{Introduction}  
\label{sec:intro}

It is commonly believed that an inflation predicts a spectrum of density perturbations similar to a power law or which can be approximated by a power law over the wide range of scales up to the current horizon value. Indeed, a smooth spectrum with barely discernible deviation is a common feature of many inflationary models based on the single scalar field theory. For example, the density perturbation spectrum in chaotic inflation with the single massive scalar field looks like 
$$
q_k\simeq\frac m\pi\ln\left[\frac{k_1}k\left(\ln\frac{ k_1}k\right)^{1/3}\right]
$$
in terms of $q$-scalar (see \cite{lukash1980}, 
\cite{bookrus}), where $k$ is a wave number, $k_1$ is the wave number at the end of inflation, and $m$ is a mass of inflaton. The running parameter of this spectrum cannot be detected at the available accuracy measurements.

During the inflationary stage of the Universe evolution, processes that disrupt the smooth spectrum of density perturbations could have taken place. Many such models have been proposed in the context of inflation with two fields, resonant amplification with oscillatory features, and others (see \cite{2023JCAP...04..011I}, numerous references wherein, {and \cite{2006PhRvD..74d3525K, 2008JCAP...06..024S, 2018PhRvD..97b3501B, 2020JCAP...08..001B, 2022JCAP...06..007I, 2020JCAP...06..013C, 2020PhRvD.102j3527Z, 2020JCAP...04..007M}}). 
The motivation for this research has been greatly enhanced by recent observational arguments in favour of primordial black holes as viable dark matter candidates. These arguments are the detection of gravitational waves from 
{merging} black holes (see \cite{ligo}) and 
{no detection of dark matter particle candidates of the Standard Model of particle physics}
(see \cite{Aleksandrov:2021}, {\cite{Nevzorov:2023}}).   
To generate primordial black holes it is necessary to amplify the density perturbation spectrum at some scales, 
{which determines the mass spectrum of black holes. Possible mechanisms of the amplification and matter power spectra can be found in aforementioned references.}
Unless one assumes the fine-tuning of related inflationary quantities, it follows that a peak-like features (also called a bump) can appear at other scales too (see, e.g. Fig.~2 of \cite{Ivanov94}). 
{If a spectral feature appears on the scales $k<10^3$~h/Mpc, it will affect nonlinear large structure formation, in particular, abundance of galaxies, their mass growth histories and clustering. The exact shape of the spectral feature depends on the inflation model, dozens of examples can be found in the references listed above. Instead of analysing nonlinear structure formation for every particular variant of such a peak, we assume that a single general shape can reproduce the most significant consequences of the introduction of the peak. We choose a Gaussian as such a general shape. In the Appendix A we demonstrate how our results could change if one uses the same method to analyse a particular example of a spectral feature predicted by one of the inflation models.}
{It is important to remark, that the location(s) of the break(s) in the inflaton potential and, consequently, the peak(s) in the matter spectrum cannot be selected a priori (they are free parameters of any such a theory).}
For example, they 
{may arise} at sub-Mpc {scales} and be responsible for some unusual feature in halo mass function at low and high redshifts. This case is under consideration in this paper.

The power spectrum of density perturbations is now measured at comoving wavenumbers $k<1$~$h/$Mpc from the Cosmic Microwave Background (CMB), large scale structure of the Universe, Ly-$\alpha$ forest \citep{Chabanier19}. No deviations from the power-law primordial spectrum have been found. The deviations at smaller scales, $1<k<10^{3}$~$h/$Mpc, would affect galaxy formation. A simple and straightforward estimate using Press-Schechter (PS) formalism \citep{press} shows that addition of a peak-like feature at $k>1$~$h/$Mpc will result in amplification of the number density of galaxy or dwarf-sized haloes at high redshifts, while this amplification will be almost smoothed away at redshift $z=0$\footnote{Figure \ref{fig:sh_tor} demonstrates this behaviour using the ST approximation.}. This means that it is difficult to find a trace of the bump in the power spectrum at $z=0$, and observations at the Epoch of Reionization (EoR) are more suitable for that.

Recently surprisingly high number of massive galaxies at $z\geq 9$ has been discovered in \textit{James Webb Space Telescope} (JWST) data \citep{Naidu2022b,Castellano22,Finkelstein22,Donnan23,Labbe23}. It is actively debated now, whether these findings are in tension with the abundance of haloes predicted by the $\Lambda$CDM \citep{Boylan-Kolchin23,Lovell22,Chen23,Prada23,Shen23}. {A number of modifications of the cosmological model have been proposed, including modifications of the power spectrum with a tilt at blue end \citep{Parashari23,Hutsi23} and a bump \citep{Padmanabhan_2023}.} Our aim is to present the expected deviations from the $\Lambda$CDM halo abundance for cosmological models with the bump in the power spectrum. We demonstrate the use of bumpy power spectrum models by applying the Extreme Value Statistics (EVS) \citep{Gumbel58} to compare observations with theoretical predictions. Recently \cite{Lovell22} have shown using EVS that JWST observations can be explained by $\Lambda$CDM mass function only if one adopts very efficient star formation at the EoR: about 100\% of baryons in haloes must transform into stars.

The halo mass function, which is needed to construct the EVS, can be obtained with modifications of the PS formalism, e.g. the more precise Sheth-Tormen (ST) formula \citep{Sheth99} but there is a couple of possible caveats here.
As shown by \cite{Klypin_2011}, ST approximation gives
a fairly accurate fit for z = 0 for a wide range of halo masses, however, it tends to overpredict the abundance of haloes at higher redshifts. This downside can be at least partially circumvented by adjusting the radius $R$ of the real space spherical top-hat window function $\hat{W}(kR)$ in the ST formula that we used 
 (for more accurate improvements over the ST approximation see e.g. \cite{Tinker_2008}, \cite{Behroozi_2013} {and \cite{Despali_2016}}).
Furthermore, while ST is known to reproduce the $\Lambda$CDM mass function quite well \citep{Jenkins2001,Reed2003}, it is not so good for cosmological models with more complicated spectra, e.g. Warm Dark Matter which has a sharp cut-off at small scales \citep{Angulo13}. So, before using ST or any other theoretical or empirical mass function for calculating the number density of haloes in cosmological models with bumpy spectra, we need to check that this technique is applicable. For this, we run a set of N-body cosmological simulations with different bumpy spectra, as well as a standard $\Lambda$CDM spectrum, and compare the simulated mass functions with the theoretical one.

Some investigations of halo formation in cosmological models with bumpy power spectra have been done in the literature. Authors of \cite{Knebe01} have analysed the impact of a positive or negative bump on the galaxy cluster mass function. They have found that PS formalism predicts the mass function quite well, but they have simulated different part of the power spectrum, $k<1$~$h/$Mpc. Since the slope of the power spectrum transfer function changes with $k$ and become more steep at higher $k$, it is not clear whether the results of \cite{Knebe01} are applicable for the range of $k$ we are interested in. Later \cite{Bagla09} have simulated a model with the bump at $k\approx 1$~$h/$Mpc, but the simulation resolution was not very high and the change of the mass function they found is moderate in comparison with the noise.
 
The paper is organised as follows:
in Section \ref{sec:cosmol} we describe the considered models;
in Section \ref{sec:simulations} we describe the details of our numerical simulations and provide the list of parameters used to run the N-body simulations.
In Section~\ref{sec:hmf} we present the ST mass functions 
adopted for modified matter spectra.
In Section \ref{sec:evs} we discuss the early halo formation realising in considered cosmological models.
Finally, we leave the Section \ref{sec:conclusions} for discussion of various implications.

\begin{figure}
    \centering
    \includegraphics[width=0.47\textwidth]{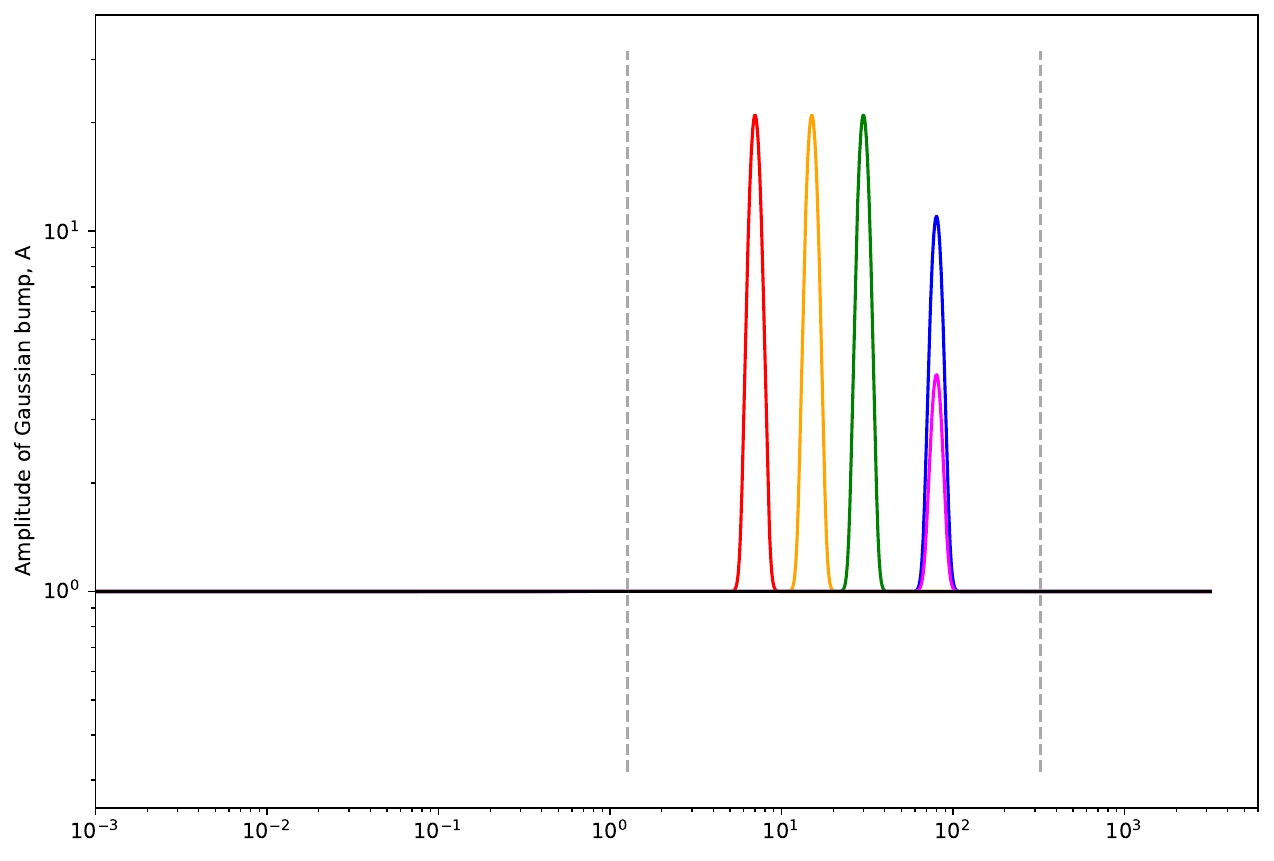}
    \includegraphics[width=0.47\textwidth]{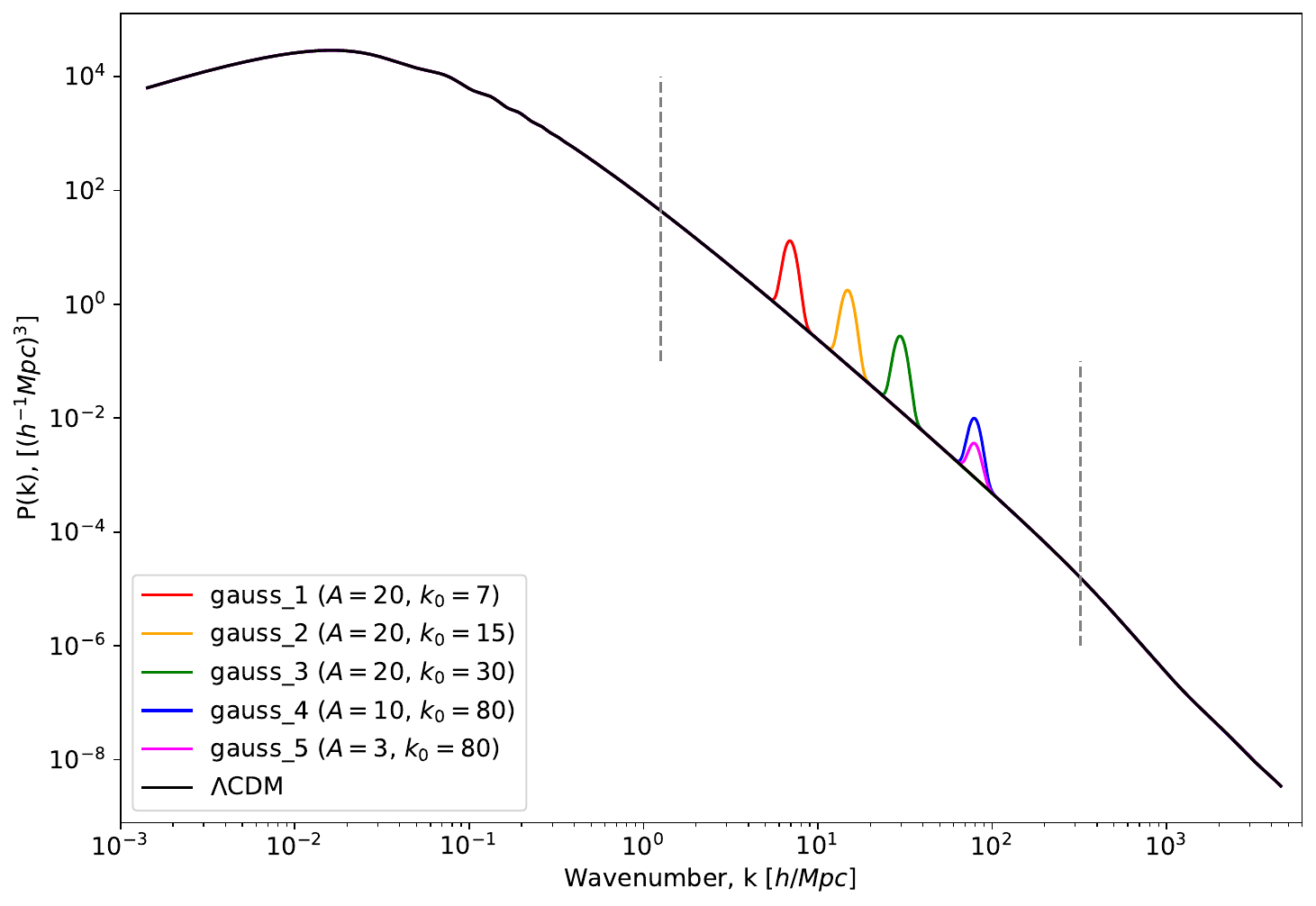}
    \caption{\textit{Top panel}: Gaussian transform functions, applied to the $\Lambda CDM$ spectrum. 
    \textit{Bottom panel:} resulting matter power spectra used in our simulation suite. \\
    On both panels grey dashed lines represent the Nyquist frequency for $(5\, Mpc/h)^3$ cube and ${512}^3$ particles resolution. The parameters of each individual spectrum can be found in the Table~\ref{tab:sim}.
    }
    \label{fig:spectra}
\end{figure}

\section{Cosmological model}
\label{sec:cosmol}
To study the influence of the peak-like features of density perturbations spectrum on the evolution of dark matter haloes we focused on the power spectra, which can be defined as a product of the standard $\Lambda$CDM spectrum and a Gaussian bump of the following form: 
\begin{equation}
    1 + A \cdot \exp \left( -\frac{(\log(k)-\log(k_0))^2}{\sigma_k^2} \right), 
    \label{bumps}
\end{equation}
where $k$ is a wave number, $A$, $k_0$, and $\sigma_k$ are bump parameters. We assume an equal value of $\sigma_k=0.1$ in all models. Numerical values of the parameters can be found in the Table~\ref{tab:sim}, whereas the Fig.~\ref{fig:spectra} illustrates the matter spectra. One can note that the shape (\ref{bumps}) is not predicted directly by simple modifications of the inflation model (see, e.g., \cite{2023JCAP...04..011I}). We consider the Gaussian shape as a universal approximation of the peak shape in various models.

The width of the peak, $\sigma_k$, does not play an important role in our studies: it is easy to show using theoretical halo mass functions (e.g., PS or ST) that the mass function does not change when $A\sigma_k = const$ for $\sigma_k \ll 1$. So there is enough to vary only one parameter, in our case the amplitude $A$. We do not consider bumps with $\sigma_k\gtrsim 1$ as this would made the study more complicated.

\begin{table*}
\caption{Most relevant parameters of the simulation suite.} 
\centering
\begin{tabular}{llllllll}
\hline
Main suite:              & $\Lambda CDM$ & $gauss\_1$ & $gauss\_2$ & $gauss\_3$ & $gauss\_4$ & $gauss\_5$     \\ \hline
Box size $(Mpc/h)$       & 5.0           & 5.0        & 5.0        & 5.0        & 5.0        & 5.0            \\
N. particles $N_{total}$ & $512^3$       & $512^3$    & $512^3$    & $512^3$    & $512^3$    & $512^3$        \\
Initial redshift         & $300$         & $10^3$     & $10^3$     & $10^3$     & $10^3$     & $10^3$         \\
Final redshift           & $8$           & $8$        & $8$        & $8$        & $8$        & $8$         \\
$k_0$                    & --            & 7          & 15         & 30         & 80         & 80             \\
$A$                      & 0             & 20         & 20         & 20         & 10         & 3              \\
$\sigma$                 & --            & 0.1        & 0.1        & 0.1        & 0.1        & 0.1            \\ 
\hline
\end{tabular}
\label{tab:sim}
\end{table*}

All simulations share the same cosmological parameters in agreement with  the values obtained by the \citet{planck}, i.e.
$\Omega_m=0.31$, $\Omega_{\Lambda}=1-\Omega_m=0.69$,
$\Omega_b=0.048$, $h=0.67$, $n_s=0.96$.


As one can see from the Table~\ref{tab:sim}, all modified spectra from \textit{gauss\_1} to \textit{gauss\_4} have Gaussian bump maxima at different values of $k_0$. The last spectrum, \textit{gauss\_5}, is the exception, and has the same $k_0$ as \textit{gauss\_4}, but smaller amplitude $A$. This was done to study the case, where the amplitude of the Gaussian bump has small impact on the resulting halo mass function.

\section{N-body Simulations}  
\label{sec:simulations}

We have run a series of 6 dark matter only simulations in a box size of $(5\,Mpc/h)^3$, $512^3$ particles each.
5 of simulations correspond to different modified matter power spectra, and one corresponds to the standard $\Lambda CDM$ one.

For our simulation suite we use the publicly available N-body code \texttt{GADGET-2}\footnote{http://wwwmpa.mpa-garching.mpg.de/~volker/gadget/} \citep{gadget}, which is widely used for cosmological simulations. This code uses a combined Tree + Particle Mesh (TreeMP) algorithm to estimate the gravitational accelerations for each particle by decomposing the gravitational forces into a long range term, computed from Particle-Mesh methods, and short scale interactions from the nearest neighbours using Tree methods, and can be used with periodic boundary conditions in the comoving frame. The code is designed to be MPI parallel, enabling it to efficiently utilise distributed computing resources by dividing the computational tasks among multiple processors, resulting in faster execution and scalability (i.e. $\cal{O}$$(N\log N)$), so it can handle a large number of particles with reasonable computational resources. 

All simulations with modified matter spectra start at $z = 1000$ in order to account for potential early formation of virialised structures,  while the simulation with $\Lambda CDM$ spectrum starts at $z = 300$. The final redshift of the simulations was set to $z = 8$, in order to reduce possible artefacts due to space periodicity of initial conditions in the relatively small-sized box.

We generate initial conditions for our simulations using publicly available code \texttt{ginnungagap}\footnote{https://github.com/ginnungagapgroup/ginnungagap}, the matter power spectrum is defined for each simulation individually by applying the transform function (\ref{bumps}), and where $\Lambda CDM$ power spectrum is generated with publicly available code CLASS \citep{CLASS}. {We use the same initial random seed number for all our simulations, so they differ only by the amplitude of the power spectrum. When the initial conditions were generated, the amplitude of the longest wavelength mode is within 20\% of the theoretical one, so the cosmic variance does not significantly affects the high-mass end of our mass functions.}

A total of 100 snapshots for each simulation were stored at redshift intervals equally-spaced in logarithmic scale, starting from $z = 35$ to $z = 8$. Halo analysis was performed with publicly available code \texttt{AHF}\footnote{http://popia.ft.uam.es/AHF/} \citep{AHF} with the assumption that each halo consists of no less than 50 particles, {while virial overdensity criterion was assumed 200 $\rho_{crit}$.}

\section{Halo mass function}   
\label{sec:hmf}
For each simulation we plotted a halo mass function (HMF) at two different redshifts: $z \simeq 13$ and $z \simeq 8$, where the former was chosen as a typical high-redshift value for the most distant galaxies found today.
While $z \simeq 8$ is the final redshift, up to which our simulation was calculated.

\begin{figure*}
    \centering
    \includegraphics[width=0.95\textwidth]{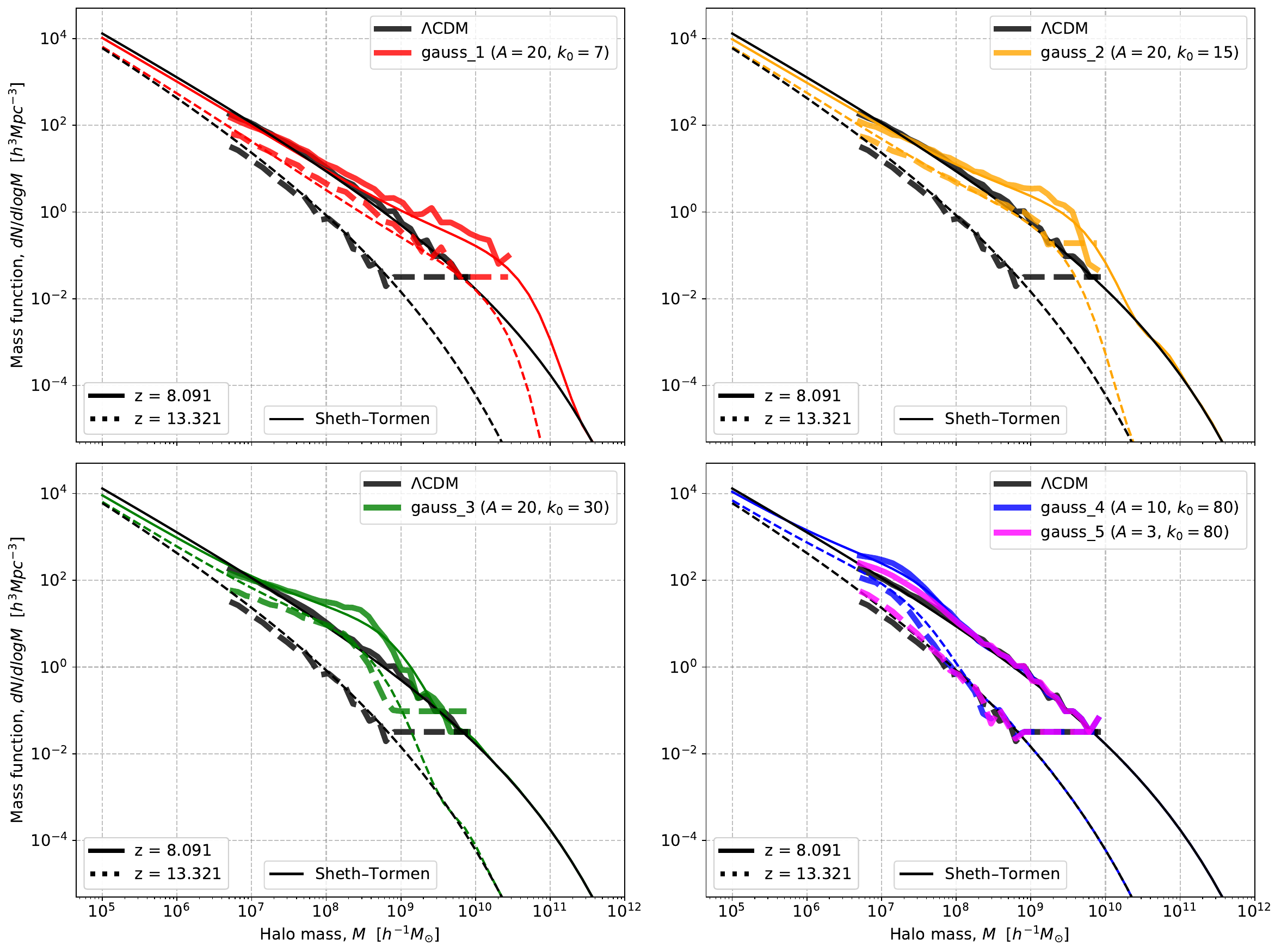}
    \caption{Halo mass function (HMF) evolution from $z \simeq 13$ (\textit{dashed} lines) to $z \simeq 8$ (\textit{solid} lines) for varying primordial matter power spectra (see the legend in the upper right corner in each panel from left to right, and from top to bottom and the Tab.~\ref{tab:sim}). \textit{Thick solid} and \textit{thick dashed} lines represent the HMF obtained from simulations, while \textit{thin solid} and \textit{thin dashed} lines represent the HMF obtained from ST approximation {(the size of the error bars for halo counts is not depicted, but can be calculated from the Poisson distribution).}}
    \label{fig:hmf}
\end{figure*}

The comparison of simulated mass functions with the theoretical (ST) ones is shown in Fig.~\ref{fig:hmf}. Each panel (from left to right and from top to bottom) includes HMFs for $\Lambda CDM$ (\textit{black} lines) spectrum and one of the modified spectra (see legend for the specific colour). \textit{Solid} lines represent HMF for $z \simeq 8$, while \textit{dashed} lines do HMF with $z \simeq 13$. As expected, an increase in the power spectrum amplitude at certain wave numbers results in a corresponding increase in the abundance of haloes with masses associated with those wave numbers, as demonstrated by the Fig.~\ref{fig:hmf}.

In addition, each panel includes the corresponding HMFs, obtained from ST approximation\footnote{However, theoretical HMF with the \textit{gauss\_5} spectrum is not included in the right low panel of the Fig.~\ref{fig:hmf} to avoid clutter, and because it differs very slightly from \textit{gauss\_4}.}.
One can see from Fig.~\ref{fig:hmf} that HMFs obtained from simulations with modified power spectrum appear to be in good accordance with the HMFs, obtained via ST approximation for the corresponding spectra. We compute ST HMF using the real space spherical top-hat window function. We have checked that it gives better coincidence with simulated HMFs than using Fourier-space top-hat or Gaussian window function for the bumpy power spectra. 

\begin{figure}
    \centering
    \includegraphics[width=0.47\textwidth]{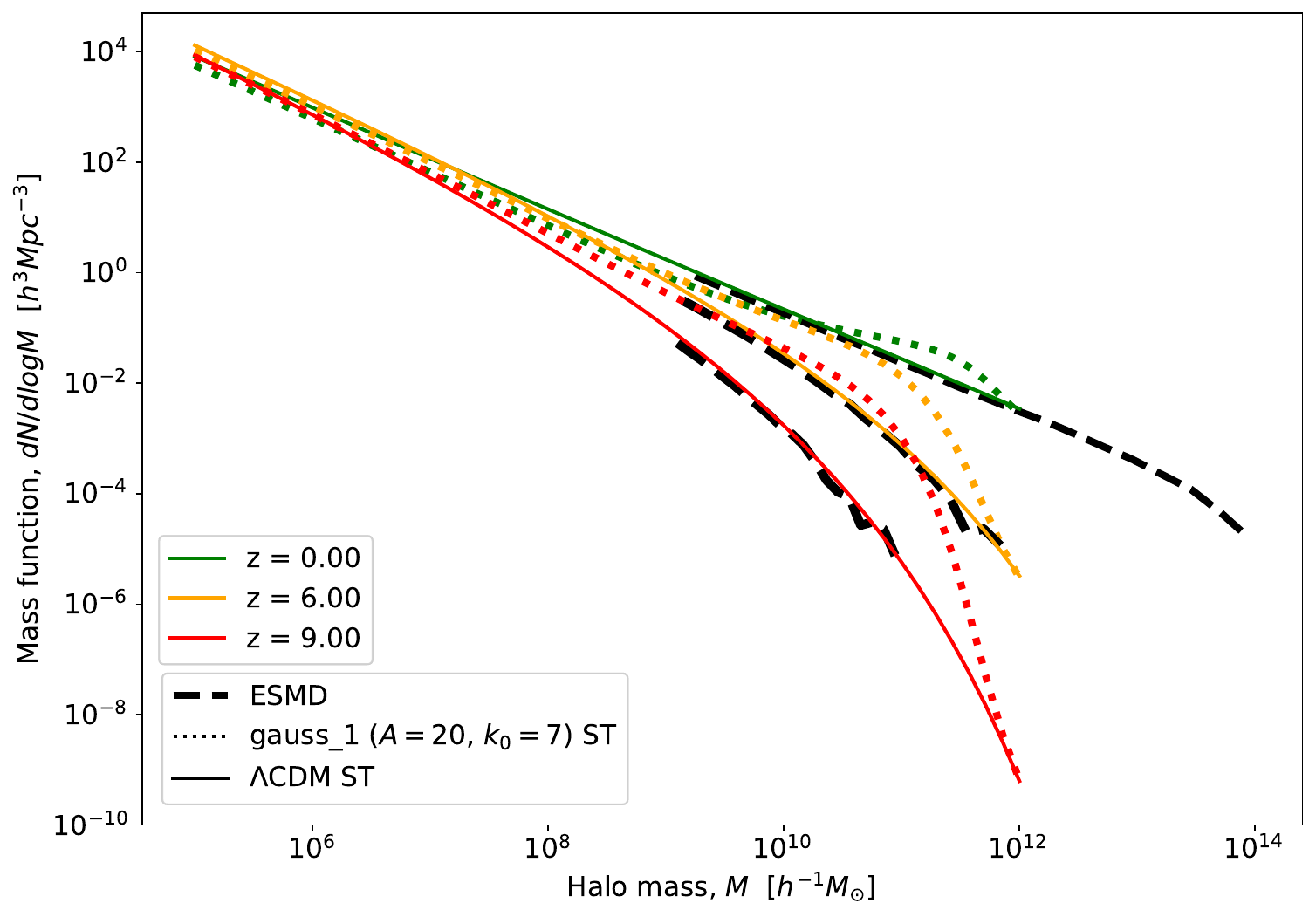}
    \caption{ST HMF for $\Lambda CDM$ spectrum (\textit{solid} line) and modified spectrum $gauss\_1$ (\textit{dotted} line) at various redshifts: \textit{red} -- $z = 9$, \textit{orange} -- $z = 6$, and \textit{green} -- $z = 0$. \textit{Thick dashed black} line corresponds to the HMF from \texttt{ESMD} simulation.}
    \label{fig:sh_tor}
\end{figure}

Considering that our simulations final redshift is $z = 8$, in Fig.~\ref{fig:sh_tor} we also compare the theoretical (ST) HMFs with the standard $\Lambda$CDM spectrum and the modified spectrum $gauss\_1$ at redshifts $z = 9$, 6, and 0. As one can see, in case of $z=6$ and $z=0$ lines, as redshift decreases, the difference between HMFs for $\Lambda CDM$ spectrum and HMFs for modified spectrum diminishes. This trend suggests that the impact of the modification on the HMF becomes less significant as the Universe evolves to lower redshifts.

Additionally, we had some concerns regarding the statistical adequacy of our simulation suite due to a relatively small size of simulation boxes $(5\,Mpc/h)^3$. Thus, we considered several snapshots from the simulation Extremely Small MultiDark Planck (\texttt{ESMD}) carried out by an international consortium within the framework of the MultiDark project\footnote{http://www.multidark.es}, and available at CosmoSim database\footnote{http://www.cosmosim.org}.
\texttt{ESMD} has a much larger box size, $(64 Mpc/h)^3$. As one can see, HMFs derived from the \texttt{ESMD} simulation, represented by the \textit{dashed black} line on the Fig.~\ref{fig:sh_tor}, closely aligns with the theoretical ST HMFs for the $\Lambda$CDM spectrum, which indicates that ST HMFs that we plotted does not noticeably overestimate the number of haloes in the case of our simulation suite as well.

Overall, the close agreement between the ST~HMFs and the simulated data in case of both $\Lambda$CDM spectrum and the modified spectra validates the utility of the ST approximation in capturing the halo mass distribution in cosmological simulations with modified spectra.

\section{Extreme value statistics with bumpy power spectra} \label{sec:evs}
Within this Section, we discuss how the effects of modified cosmological power spectrum might affect the observations of halo formation in the early Universe. We test how well different power spectra fit the observational data on $z>8$ galaxies with EVS approach. The idea behind the EVS is to test if the most extreme observed value of some property is indeed an outlier. In our case, we test the maximal halo mass of galaxies observed by JWST at high redshifts. To infer halo masses from observations, we use the stellar fraction $f_*$ and we find for each observed galaxy the minimal value of $f_*$ for which this galaxy is not a $3\sigma$ outlier from the theoretical maximal halo mass distribution.

\begin{figure*}
\centering
\includegraphics[width=0.99\textwidth]{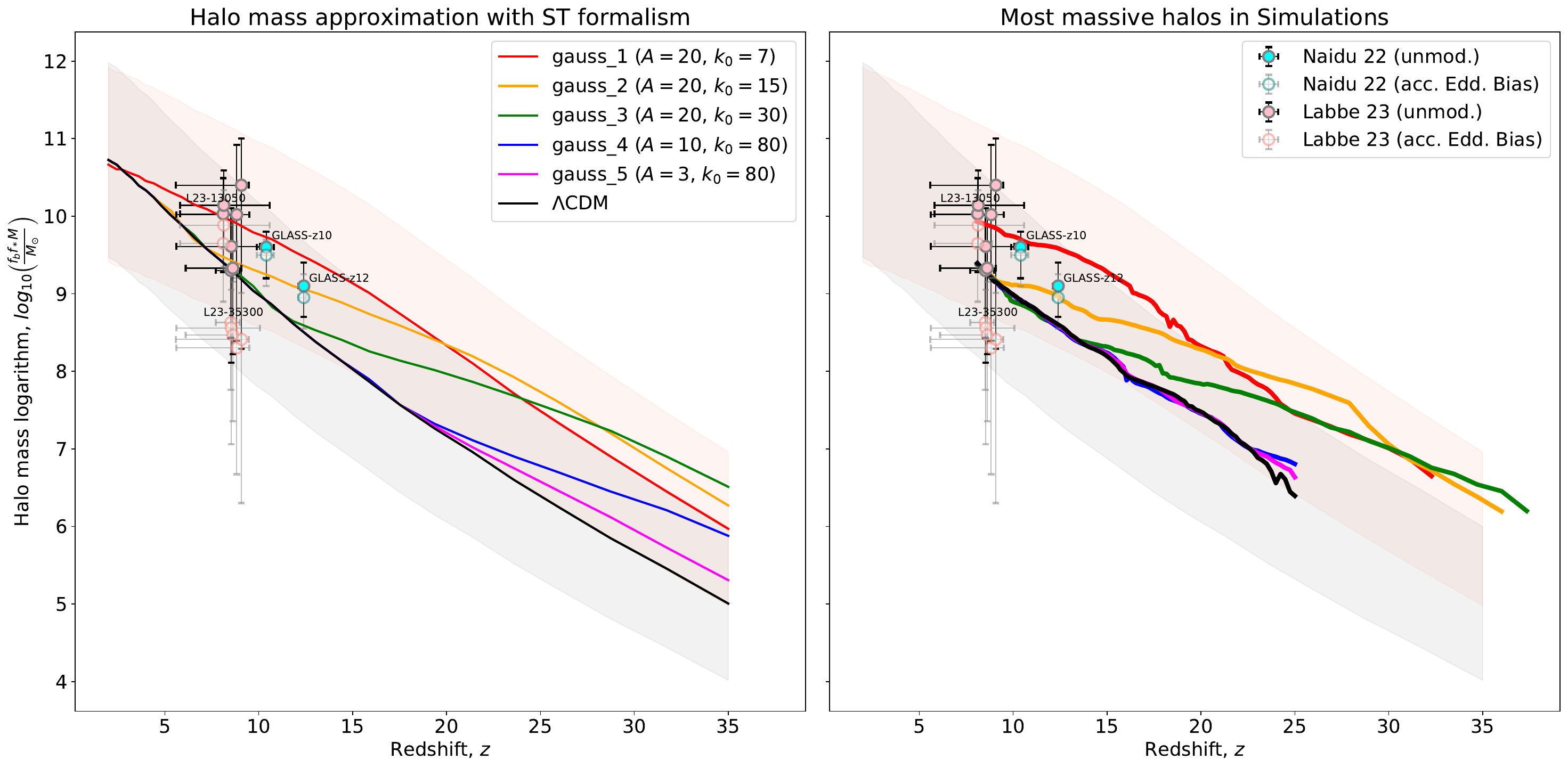}
\includegraphics[width=0.99\textwidth]{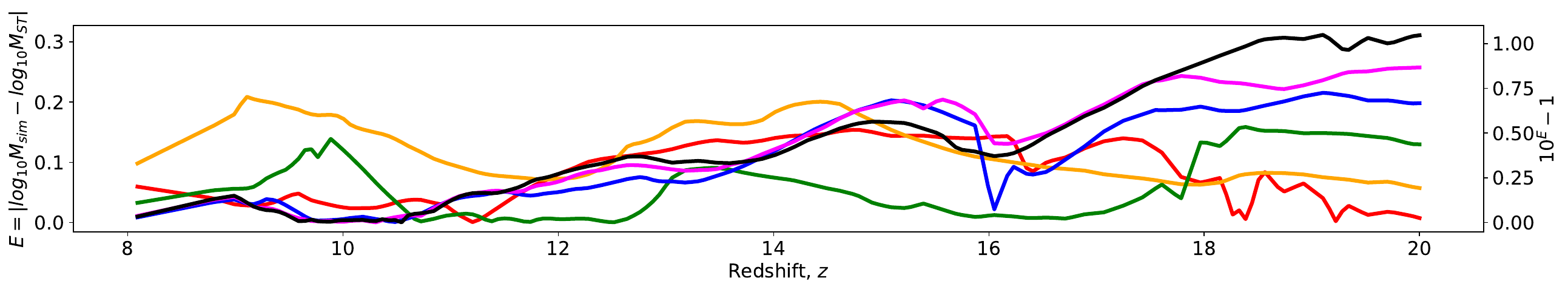}
\caption{The mass of the most massive halo in the simulation cube, multiplied by the fixed baryon fraction $f_b = 0.157$, as a function of redshift $z$ for varying primordial matter power spectra (see the legend in the upper right corners).\\
\textit{Left} panel presents an approximation of the masses of the most massive haloes using the theoretical (ST) mass function, depicted with \textit{thin solid} lines.\\
\textit{Right} panel depicts the masses of the most massive halo in our numerical simulations at a given redshift, represented by the \textit{thick solid} lines.\\
The observational data from Table~\ref{tab:sources} are shown as filled \textit{cyan} and \textit{pink} dots with error bars.\\
The filled \textit{light-grey} {and \textit{light-red}} areas represent the $3\sigma$ ranges {for $\Lambda CDM$ and $gauss\_1$ mass approximations} derived from EVS, and {they are} depicted in both the \textit{left} and \textit{right} panels.\\
\textit{Bottom} panel represents the difference between the simulation and ST masses respectively.}
\label{fig:mass_vs_z}
\end{figure*}

The data we use are presented in the Table~\ref{tab:sources} and also displayed on the Fig.~\ref{fig:mass_vs_z}. Filled \textit{cyan} dots represent the last two sources from Table~\ref{tab:sources}, while filled \textit{pink} dots represent the rest of the sources from the Table~\ref{tab:sources}. When comparing observations to the theory, one needs to take into account the Eddington Bias \citep{Eddington13}. Due to the steepness of the HMF and galaxy stellar mass function, measurements of low mass objects can be easily upscattered. White-space dots with coloured border represent masses, corrected for Eddington Bias as follows:
\begin{align}
\ln M_{\mathrm{Edd}}=\ln M_{\mathrm{obs}}+\frac{1}{2} \epsilon \sigma_{\ln M}^2 \text {, }
\end{align}
where $\epsilon$ is the local slope of the underlying halo mass function, and $\sigma_{\ln M}$ is the uncertainty in the halo/stellar mass estimate.

For the given area of the survey we compute the PDF of the most massive halo as a function of $z$ for $\cal{N}$ observations (see \cite{Lovell22}) as:
\begin{align}
\Phi(M_{max}=m, \mathcal{N}) = \mathcal{N} f(m) \left[ F \right]^{\mathcal{N}-1}\,,
\label{eqn:evs}
\end{align}
where $f(m)$ and $F(m)$ for a fixed fraction of the sky $f_{sky}$ between redshifts $z_{\min}$ and $z_{\max}$ are calculated as follows:
\begin{align}
f(m) &= \frac{f_{\text{sky}}}{n_{\text{tot}}} \left[ \int_{z_{\min}}^{z_{\max}} dz \frac{dV}{dz} \frac{dn(m, z)}{dm} \right]\,, \\
F(m) &= \frac{f_{\text{sky}}}{n_{\text{tot}}} \left[ \int_{z_{\min}}^{z_{\max}} \int_{-\infty}^{m} dz \,dM \frac{dV}{dz} \frac{dn(M, z)}{dM} \right]\,,
\end{align}
where $n_{tot}$ is a normalisation factor giving the total number density of haloes:
\begin{align}
n_{\mathrm{tot}}=f_{\mathrm{sky}}\left[\int_{z_{\min}}^{z_{\max}} \int_{-\infty}^{\infty} d z\, d M \frac{d V}{d z} \frac{d n(M, z)}{d M}\right]\, .
\end{align}
Using the aforementioned equations along with the 
Eq.~(\ref{eqn:evs}), while considering an observational survey volume with an area of approximately 40 arcmin$^2$ as reported in \cite{Naidu2022a, Naidu2022b, Labbe23} one can find a EVS halo for a given survey. We plot the median of the PDF for $\mathcal{N}=1$, as well as the $3\sigma$ interval around the median in the \textit{left} panel of Fig.~\ref{fig:mass_vs_z}.

The thin solid coloured lines on the \textit{left} panel of Fig.~\ref{fig:mass_vs_z} correspond to the ST approximation, while the thick solid lines on the \textit{right} panel represent the most massive halo's mass in our numerical simulations for a given redshift. The lines are colour-coded in the same manner as in Fig.~\ref{fig:spectra} and Fig.~\ref{fig:hmf}. The $3\sigma$ ranges for $\Lambda CDM$ and $gauss\_1$ mass approximations, obtained from EVS statistics calculations, are depicted as filled \textit{light-grey} and \textit{light-red} areas respectively and depicted in both \textit{left} and \textit{right} panels of the Fig.~\ref{fig:mass_vs_z}. The $3\sigma$ range for other models has similar width as for the $\Lambda$CDM model, i.e. from approximately $1/20$ to $20$ of the median. Additionally, both the masses approximated from theoretical HMF at the \textit{left} panel and the halo masses from simulations at the \textit{right} panel were adjusted by the baryon fraction, $f_b = 0.157$, 
and the stellar fraction $f_*$. 
An upper limit on this fraction, $f_*=1$, is equivalent to the assumption that all the baryons are converted into stars. Observations and theoretical models show that the distribution of stellar fraction peaks at $f_*<0.2$ at $z<10$, see, e.g. \citep{Giodini_2009,Silk2018,Lovell22}. On the other hand, the observational data at pre-reionization epoch ($z>10$) is missing, while some theoretical models predict high star formation efficiencies, up to $f_*=0.8$ \citep{Susa_2004}.

To elucidate the relationship between the bump and the value of $f_*$, we calculate the values of this quality in $\Lambda$CDM and $gauss\_1$ for each of high-z sources. The Table~\ref{tab:sources} shows the minimal values of $f_*$ for $\Lambda$CDM and $gauss\_1$ respectively, which are required for the given galaxy to fall into $3\sigma$ range. As can be seen, with the assumption of the bumpy power spectrum, most of the galaxies need several times smaller value of $f_*$.

As can be seen from Fig.~\ref{fig:mass_vs_z}, the masses of the most massive haloes in the simulations with modified spectra differ significantly from those in the $\Lambda$CDM model above some redshift, which depends on the position of the bump. At smaller redshifts this quantity converges to the $\Lambda$CDM model. In particular, our model $gauss\_1$ fits the observations much better than the $\Lambda$CDM, allowing smaller, more reasonable values of the stellar fraction in high-z galaxies.

Both masses approximated from ST and the halo masses from simulations with modified spectra exhibit a similar trend with the deviations less than 30\% for the majority of spectra up to $z \approx 12$, although some discrepancy might be observed at higher redshift values (see the \textit{bottom} panel of Fig \ref{fig:mass_vs_z}). This implies that ST approximation can be used for calculating the EVS for bumpy power spectra.\\
\begin{table*}
\centering
\caption{Comparison of several observed galaxy candidates with $\Lambda$CDM simulations, sorted by Redshift (z).}
\begin{tabular}{|l|l|c|c|c|l}
\hline
Source ID  & Redshift (z)         &  log$_{10}$($M_*$/\msun) & {$\Lambda$CDM $f_{*\,min}$} & {$gauss\_1$ $f_{*\,min}$} & Refs. \\
\hline
id-2859    & $8.11^{+0.75}_{-2.30}$ & $10.03^{+0.46}_{-0.75}$ & {0.26} & {0.07} &  \cite{Labbe23}\\
id-13050   & $8.14^{+2.45}_{-2.33}$ & $10.14^{+0.45}_{-0.54}$ & {0.34} & {0.09} &  \cite{Labbe23} \\
id-16624   & $8.52^{+0.46}_{-0.80}$ & $9.30^{+0.72}_{-0.87}$  & {0.06} & {0.01} &  \cite{Labbe23} \\
id-21834   & $8.54^{+1.52}_{-2.92}$ & $9.61^{+0.49}_{-1.50}$  & {0.12} & {0.03} &  \cite{Labbe23} \\
id-39575   & $8.62^{+0.45}_{-2.51}$ & $9.33^{+0.69}_{-1.11}$  & {0.07} & {0.02} &  \cite{Labbe23} \\
id-14924   & $8.83^{+0.67}_{-3.22}$ & $10.02^{+0.90}_{-1.63}$ & {0.38} & {0.08} &  \cite{Labbe23} \\
id-35300   & $9.08^{+0.40}_{-3.50}$ & $10.40^{+0.60}_{-2.11}$ & {1.02} & {0.21} & \cite{Labbe23} \\
GLASS-z10  & $10.4^{+0.4}_{-0.5}$   & $9.6^{+0.2}_{-0.4}$     & {0.31} & {0.05} & \cite{Naidu2022b} \\
GLASS-z12  & $12.4^{+0.1}_{-0.3}$   & $9.1^{+0.3}_{-0.4}$     & {0.26} & {0.03} & \cite{Naidu2022b} \\
\hline
\end{tabular}
\label{tab:sources}
\end{table*}


\section{Discussion and conclusions}   
\label{sec:conclusions}

In the paper we studied a cosmological model with modified power spectrum of density perturbations. In this model a matter power spectrum has a Gaussian bump with a certain amplitude $A$ and a scale location $k_0$ (see eq.(\ref{bumps})). We considered five different models with parameter values summarised in the Table~\ref{tab:sim}. All models were studied with both numerical N-body simulations and analytical ST approximation. The latter supported the results of the simulations and was used to relate the observational data with our analysis using the EVS approach.

We have found that models with a bumpy spectra have higher abundance of haloes which could host first galaxies observed at $z>10$. At smaller redshifts the difference in the mass function with respect to the $\Lambda$CDM model vanishes. Models with the bumpy power spectrum can better explain the results recently obtained from the observations of JWST \citep{Naidu2022b,Labbe23}{, by allowing reasonable star formation efficiency, $f_*<0.2$. At the same time, in the case of $\Lambda$CDM power spectrum, several of the sources detected require $f_*>0.3$, and one source requires $f_*\approx 1$.} In particular, a bump with the amplitude $A=20$ and position at $k_0=7$~$h$/Mpc can explain the presence of objects with the mass $\sim10^{10}$~M$_\odot$ at $z>15$. It is possible to fine-tune the bump for the particular value of the star formation efficiency, but we haven't done this exercise because the data on halo masses, to our opinion, is yet not enough reliable to derive the exact parameters of the bump from it. Our goal was to show that the models with the bumpy spectrum in principle can explain the overabundance of haloes at high redshifts and converge to the $\Lambda$CDM at low redshifts. 

{Nowadays a number of studies propose different modifications to the power spectrum to explain the abundance of high-z galaxies \citep{Parashari23,Padmanabhan_2023,Hutsi23}. An analysis of galaxy clustering \citep{Munoz23} as well as sizes of haloes and galaxies \citep{Demianski23} could help to distinguish between different models, including $\Lambda$CDM. We plan to use our simulations of the bumpy spectra for such an analysis in future.
A bump at $k\sim7-10$~$h$/Mpc in the power spectrum can have several observational consequences.} As can be seen, e.g., from our Fig.~\ref{fig:mass_vs_z}, in the bumpy models haloes with masses below $10^{10}$~M$_\odot$ form earlier than in the $\Lambda$CDM. In particular, in the volume considered, haloes with $M=10^7$~M$_\odot$ appear at $z\approx20$ in the $\Lambda$CDM, and at $z\approx 35$ in $gauss\_1$ model. Thus, in the models with the bump the first stars would appear earlier, and there will be more time to seed and grow supermassive black holes. The next step of testing the models with the bump would be to predict the 21~cm signal from Reionization in these models (see, e.g., \citet{Munoz20}).



So, the considered class of cosmological models can be responsible for early dark matter halo formation and results in early star formation and galaxy nuclei activity. 

\section*{Acknowledgements}
Authors thank A.G. Doroshkevich and P.B. Ivanov for fruitful discussions {and the anonymous referee for their valuable remarks.}

\section*{Data availability}
The data underlying this article will be shared on reasonable request to the corresponding author.



\bibliographystyle{mnras}
\bibliography{refs} 


\appendix
{
\section{Mass function for a particular inflation model}}

\begin{figure*}
    \centering
    \includegraphics[width=0.45\textwidth]{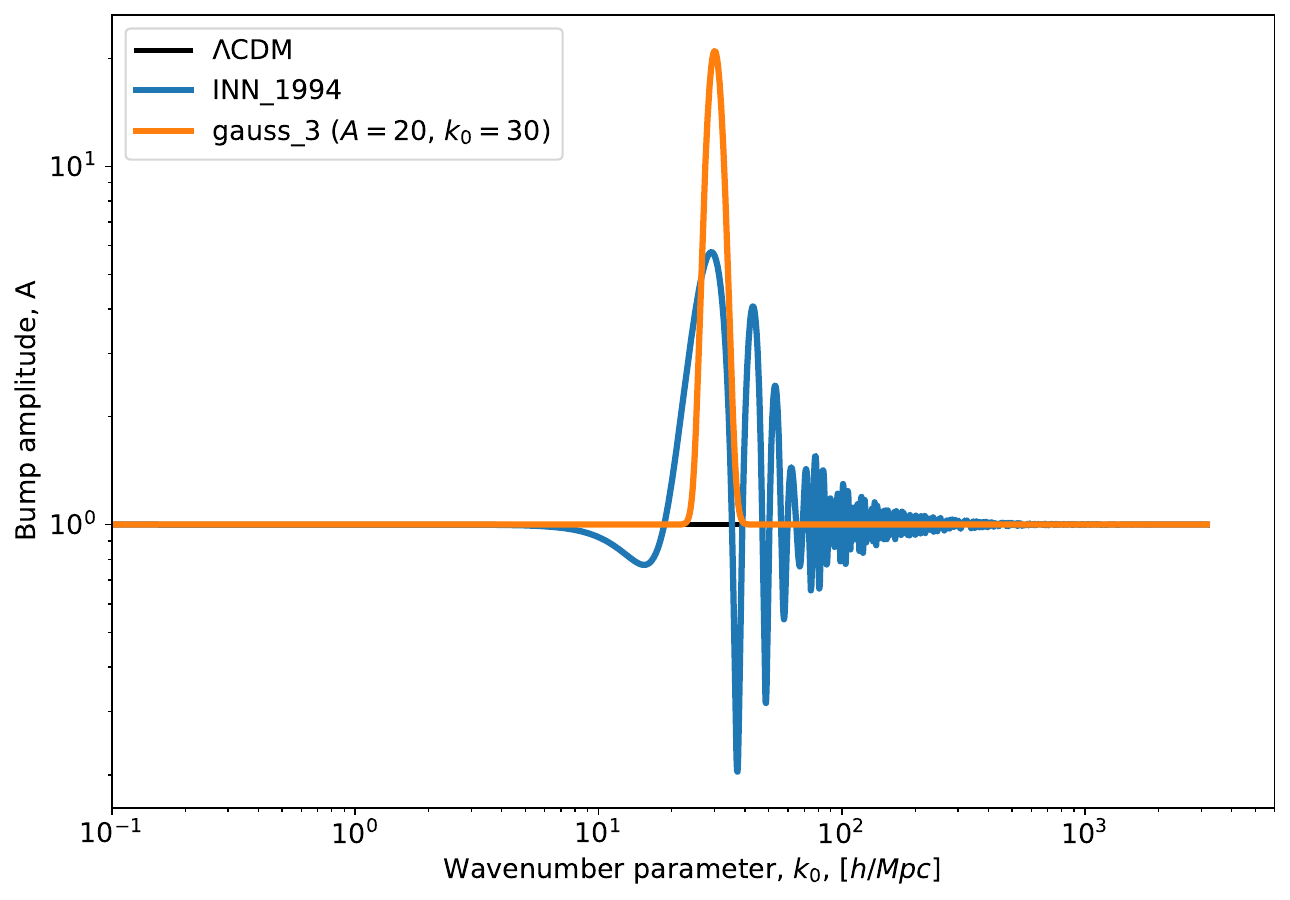}
    \includegraphics[width=0.45\textwidth]{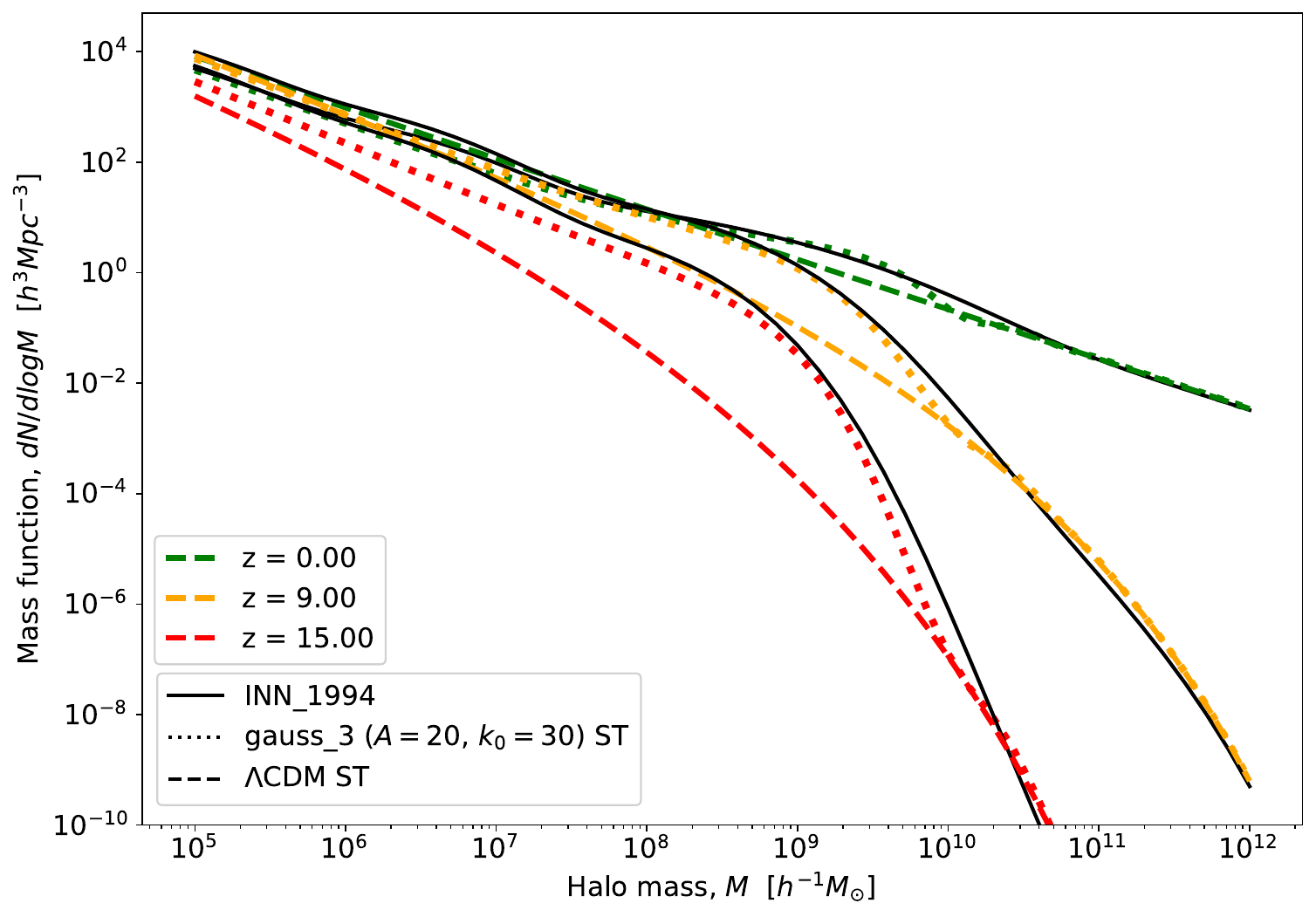}
    \caption{\textit{Left panel}: Gaussian transform function $gauss\_3$ (\textit{solid orange} line) along with $\Lambda CDM$ (\textbf{solid black} line), as well as modulation function from INN\_1994 (\textit{solid blue} line), as a function of wavenumber parameter $k$.\\
    \textit{Right panel:} resulting ST HMF for $\Lambda CDM$ spectrum (\textit{dashed} line), modified Gauss-bump spectrum $gauss\_3$ (\textit{dotted} line) and INN\_1994 (\textit{solid black} line) spectrum at various redshifts: \textit{red} -- $z = 15$, \textit{orange} -- $z = 6$, and \textit{green} -- $z = 0$.
    }
    \label{fig:INN_94}
\end{figure*}

{There is a broad class of inflationary models with a peak-like structure in the density perturbation spectrum. For example, one can consider a model whose inflaton potential $V(\phi)$ has two breaks and plateau between them, as was done in \cite{Ivanov94} (hereinafter we call this model INN\_1994). 
}
{The consideration of such a feature in the potential results in increase of the density perturbation amplitude at certain range of wavenumbers $k_2 < k < k_1$, without changing the spectrum outside this range (see Fig.~\ref{fig:INN_94}).} 

{The \textit{left} panel of Fig.~\ref{fig:INN_94} illustrates an example of the modulation function $D(x)$ describing  (\textit{solid blue} line) the quotient of the INN\_1994 model power spectrum and a standard $\Lambda$CDM. Following to notations of \cite{Ivanov94},   $D(0) = D(\infty) = 1$ and $\frac{k_1}{k_2} = 1.25$. \textit{Solid orange} line represents the Gaussian bump transform function, corresponding to $gauss\_3$ spectrum. We pick the values of the parameters for INN\_1994 model to obtain the same integral beneath the $D(x)$ as in our $gauss\_3$ bump, which should result in a similar mass function, according to the reasoning given in Section~\ref{sec:cosmol}}. 

{The \textit{right} panel of Fig.~\ref{fig:INN_94} illustrates the resulting mass functions for the given spectra (modified Gauss-bump spectrum $gauss\_3$ and INN\_1994, as well as $\Lambda CDM$ spectrum) using the ST approximation. As can be seen, the difference between the corresponding mass functions is minimal and manifests mostly in characteristic fluctuations due to periodic behaviour of the modulation function $D(x)$.}


\bsp	
\label{lastpage}
\end{document}